\documentclass[a4paper, 11pt]{article}
\usepackage[utf8]{inputenc}
\usepackage{graphicx}
\usepackage{subfigure}
\usepackage{url}
\usepackage{siunitx}
\usepackage[hidelinks]{hyperref}
\usepackage{authblk}
\usepackage{amsfonts}

\begin{document}
\title{Crowdbreaks: Tracking Health Trends using Public Social Media Data and Crowdsourcing}
\author[1]{Martin Mueller \footnote{martin.muller@epfl.ch}}
\author[1]{Marcel Salathé \footnote{marcel.salathe@epfl.ch}}

\affil[1]{Global Health Institute, School of Life Sciences, EPFL, Lausanne, Switzerland}

\maketitle

\begin{abstract}
  In the past decade, tracking health trends using social media data has shown great promise, due to a powerful combination of massive adoption of social media around the world, and increasingly potent hardware and software that enables us to work with these new big data streams.
  At the same time, many challenging problems have been identified.
  First, there is often a mismatch between how rapidly online data can change, and how rapidly algorithms are updated, which means that there is limited reusability for algorithms trained on past data as their performance decreases over time.
  Second, much of the work is focusing on specific issues during a specific past period in time, even though public health institutions would need flexible tools to assess multiple evolving situations in real time.
  Third, most tools providing such capabilities are proprietary systems with little algorithmic or data transparency, and thus little buy-in from the global public health and research community. 
  Here, we introduce Crowdbreaks, an open platform which allows tracking of health trends by making use of continuous crowdsourced labelling of public social media content.
  The system is built in a way which automatizes the typical workflow from data collection, filtering, labelling and training of machine learning classifiers and therefore can greatly accelerate the research process in the public health domain.
  This work introduces the technical aspects of the platform and explores its future use cases. 
\end{abstract}

\section{Introduction}
In the past years, data derived from public social media has been successfully used for capturing diverse trends about health and disease-related issues such as flu symptoms, sentiments towards vaccination, allergies, and many others~\cite{culotta2010towards,paul2011you,Salathe2011a,paul2012model,parker2013framework}.
Most of these approaches are based on natural language processing (NLP) and share a common workflow. 
This workflow involves data collection, human annotation of a subset of this data, training of a supervised classifier, and subsequent analysis of the remaining data.
The approach has proven promising in many cases, but it also shares a few shortcomings.
A major drawback of this type of research process is that a model, which was trained on data from previous years, might not generalize well into the future.
This issue, commonly known as concept drift~\cite{widmer1996learning}, may not necessarily be only related to overfitting, but may simply be a consequence of how language and content, especially on the internet, evolve over time.
A similar effect has been suggested to be the main reason for the increasing inaccuracy of Google Flu Trends (GFT), one of the most well-known flu surveillance systems in the past~\cite{ginsberg2009detecting}. 
After launching the platform in 2003, GFT's model had been retrained in 2009, which led to a significant improvement of its performance in the following years.
However, during the influenza epidemic in 2012/13, the model's performance decreased again and overestimated the extent of the epidemic by a large margin.
Shortly after, it was discontinued~\cite{Lazer2014,Butler2013}.\par

Apart from the issue of model drift, a second issue associated with current NLP models is that the collection of large amounts of labelled data, usually through platforms such as Amazon Turk, is very costly.
Labelling a random subset of the collected social media data may be inefficient, as depending on the degree of filtering applied, large fractions of the collected data are possibly not relevant to the topic, and therefore have to be discarded.\par

Lastly, there is a growing interest in the public health field to capture more fine-grained categorizations of trends, opinions or emotions. 
Such categorizations could allow to paint a more accurate picture of the nature of the health issue at hand.
However, multi-class annotations of a large sample of data again exponentially increases costs.\par

Here, we introduce Crowdbreaks\footnote{\url{https://www.crowdbreaks.org}}, a platform targeted at tackling some of these issues. 
Crowdbreaks allows the continuous labelling of public social media content in a crowdsourced way.
The system is built in a way which allows algorithms to improve as more labelled data is collected.
This work describes the functionalities of the platform at its current state as well as its possible use cases and extensions.

\section{Related Work}
In recent years, a number of platforms have been launched which allow the public to contribute to solving a specific scientific problem.
Among many others, examples of successful projects include the Zooniverse platform (formerly known as Galaxy Zoo)~\cite{Simpson2014}, Crowdcrafting~\cite{crowdcrafting}, eBird (a platform for collecting ornithological data)~\cite{Wood2011}, and FoldIt (a platform to solve protein folding structures)~\cite{Khatib2010}.
Many of these projects have shown that citizen science can be used to help solve complex scientific problems.
At the same time, there is a growing number of platforms which offer monetary compensations to workers for the fulfillment of microtasks (the most prominent example being Amazon Turk\footnote{\url{https://aws.amazon.com/}}).
These platforms gain importance as the need for large amounts of labelled data for the training of supervised machine learning algorithms increases.
Previous work focused mostly on efficiency improvement of large-scale human annotation of images, e.g.\ in the context of the ImageNet project~\cite{Russakovsky2015}. 
Most of these improvements include better ways to select \textit{which} data to annotate, \textit{how} to annotate (which is a UI specific problem) and what type of annotations (classes and subclasses) should be collected~\cite{kovashka2016crowdsourcing}.
Online task assignment algorithms have been suggested which may consider both label uncertainty as well as annotator uncertainty during the annotation process~\cite{Welinder2010,Ho2012}.
Results suggest that this allows for a more efficient training of algorithms.
More recently, a crowd-based scientific image annotation platform called Quantius has been proposed, showing decreased analysis time and cost~\cite{Hughes2017}.
To our knowledge, no similar work has been proposed with the regard to the human annotation of textual data (such as tweets).

\section{Platform overview}
Crowdbreaks is a platform which aims at automatizing the whole process from data collection (currently through Twitter), filtering, crowdsourced annotation and training of Machine Learning classifiers. 
Eventually these algorithms can help evaluate trends in health behaviours, such as vaccine hesitancy or the risk potential for disease outbreaks.\par
\begin{figure}[!ht]
\centering
\includegraphics{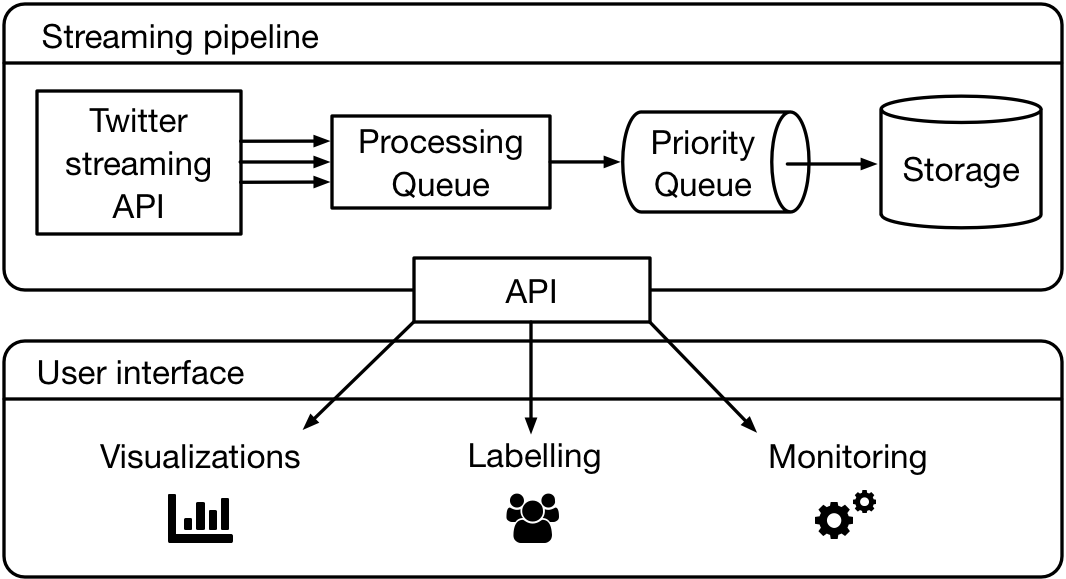}

\caption{Overview of the Crowdbreaks platform.}
  \label{fig:fig1}
\end{figure}

Crowdbreaks consists of a data collection pipeline (``streaming pipeline'') and a platform for the collection of labelled data (``user interface''), connected through an API (Application Programming Interface), as schematized in figure~\ref{fig:fig1}.

\subsection{Streaming pipeline}
\label{sec:streaming_pipeline}
Currently Crowdbreaks consumes data from the Twitter streaming API only, therefore the rest of this work will focus on tweets as the only data source.
However, it could be extended to any textual data which can be collected in the form of a data stream through an API.
The Twitter API allows for the filtering of tweets by a specific set of keywords in real-time.
Tweets collected contain at least one exact match within certain fields of the tweet object.
Incoming tweets are put on a background job queue for filtering, pre-processing, geo-tag enrichment, and annotation with metadata, such as estimated relevance or sentiment (more on this in section~\ref{sec:future_work}).
Apart from filtering by a simple list of keywords mentioned before, Crowdbreaks also allows to further filter content by applying complex keyword queries, such as \texttt{(keyword1 OR keyword2) AND keyword3}.
After these processing steps, tweets are stored in a database.
Based on a relevance score~(e.g.\ the uncertainty of a predicted label, see section~\ref{sec:training}) the tweet IDs are also pushed into a priority queue for subsequent labelling.
Once the priority queue has reached a certain size, older items with low priority are removed from the queue and replaced with more recent items.
Therefore the queue keeps a pool of recent and supposedly relevant tweets for labelling.
Once a tweet has been labelled, it is ensured that the same tweet will be labelled by a certain number of distinct users in order to reach a consensus.

\subsection{User interface}
\begin{figure}
\centering
\includegraphics[width=0.9\textwidth]{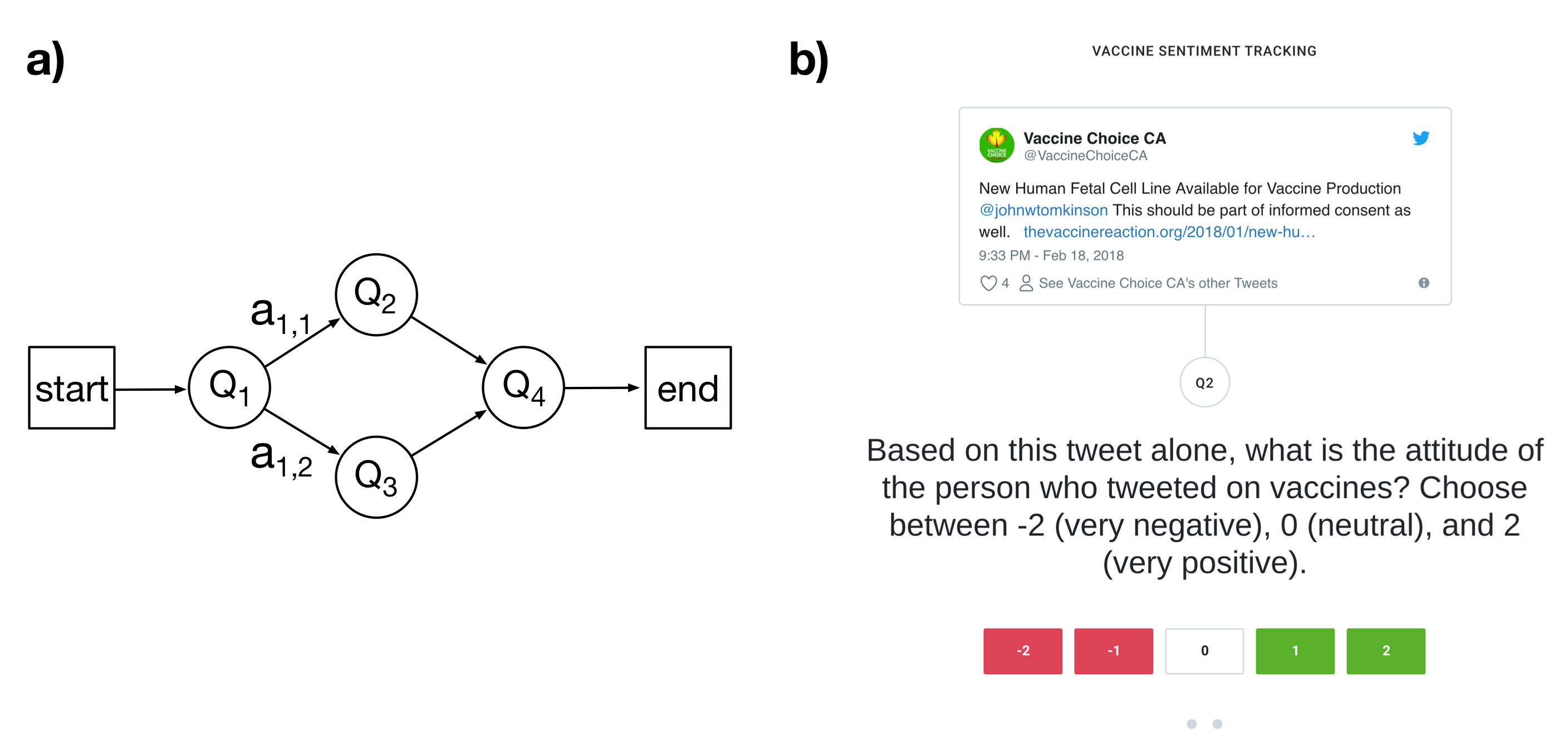}
  \caption{
    a) A very simple example of a question sequence. Questions are denoted by $Q$, answers by $a$ and the arrows designate the possible transitions between questions. 
    In the given example, different questions are reached depending on whether an annotator answers $Q_1$ with $a_{1,1}$ or $a_{1,2}$. 
    b) Screenshot of an example question for determining the vaccine sentiment of a tweet which has been deemed relevant to the topic.   
  }
  \label{fig:fig2}
\end{figure}

The user interface allows labelling of tweets based on answering of a sequence of questions.
Arbitrary question sequences can be defined, which allow the annotation of multiple classes and subclasses to a single tweet.
Most commonly, different follow-up questions would be asked depending on the answers given previously, e.g.\ whether or not the tweet is relevant to the topic at hand (see figure~\ref{fig:fig2}a). 
In the beginning of a question sequence an API call is made to the streaming pipeline to retrieve a new tweet ID from the priority queue (see section~\ref{sec:streaming_pipeline}).
Every question a user answers creates a new row in a database table, containing the respective user, tweet, question and answer IDs.
After the user has successfully finished the question sequence the respective user ID is then added to a set, in order to ensure that the same tweet is not labelled multiple times by the same user.\par
Crowdbreaks supports multiple projects, each project may be connected to its own data stream from Twitter.
New projects can be created through an admin interface, making it possible to control both the data collection, as well as to define project-specific question sequences.
Eventually, visualizations, such as sentiment trends over time, may be presented to the public user, allowing the users to see the outcomes of their work.
Crowdbreaks also features an integration of the question sequence interface with Amazon Turk, allowing the collection of labelled data through paid crowdworkers as an alternative to public users.

\subsection{Sentiment analysis}

\subsubsection{Algorithms}
\label{sec:training}
In recent years, algorithms for sentiment analysis based on word embeddings have become increasingly more popular compared to traditional approaches which rely on manual feature engineering~\cite{Bengio2003,mikolov2013efficient,joulin2016bag}.
Word embeddings give a high-dimensional vector representation of the input text, usually based on a pre-trained language model.
Although these approaches may not consistently yield better results compared to traditional approaches, they allow for an easier automatization of the training workflow and are usually more generalizable to other problems.
This is a desirable property in the context of Crowdbreaks, as it aims to further automatize this process and retrain classifiers automatically as more labelled data arrive.
Furthermore, pre-trained word embeddings based on large Twitter corpora are available in different languages, which also make them interesting for following health trends in languages other than English~\cite{deriu2017leveraging}.

\subsubsection{Active Learning}
\begin{figure}[!ht]
\centering
\includegraphics{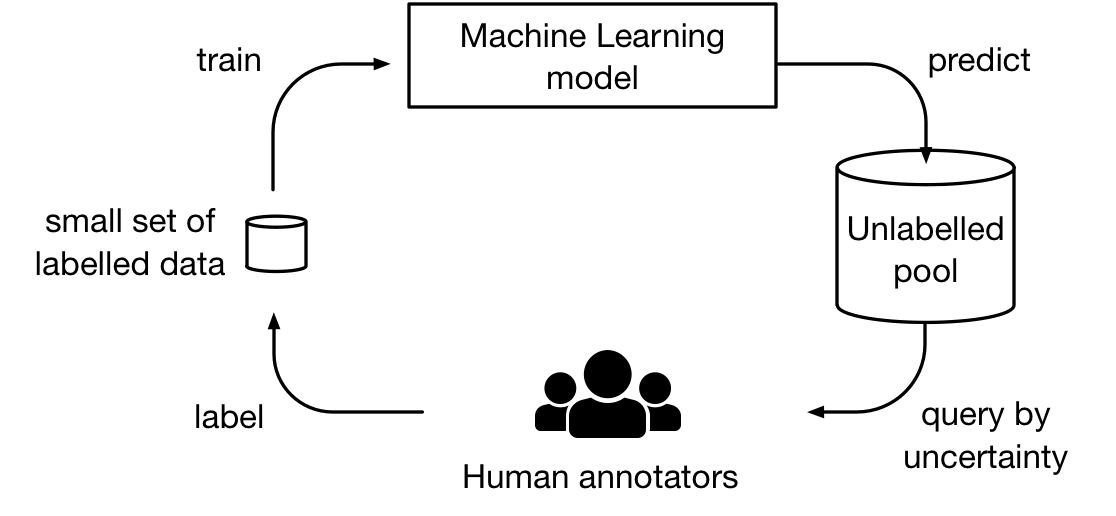}
  \caption{
    Crowdbreaks can be seen as an active learning framework which allows to improve algorithms as more labels are collected.
    In this example, an algorithm tries to learn sentiments from tweets and is given an initial small set of labelled data to be trained on. 
    This algorithm may then be used to predict the labels and label uncertainty of newly collected tweets.
    Subsequently, tweets which the algorithm is most uncertain about will be presented to human annotators.
    As new labelled data is generated, the algorithm is retrained to further improve in performance.
  }
  \label{fig:fig3}
\end{figure}

Active learning frameworks have been proposed for a more efficient training of classifiers in the context of word embeddings~\cite{Kholghi2017,Zhang2016}. 
These frameworks allow algorithms to be trained with a much smaller number of annotated data, compared to a standard supervised training workflow~(see figure~\ref{fig:fig3}).
The query strategy, which is usually related to label uncertainty, is generally the critical component for the relative performance speed-up of these methods. 
In the context of Crowdbreaks, we are not only prioritizing data with higher label uncertainty, but also data which is more recent in time.
Therefore, we are faced with a trade-off between exploration of more recent data vs.\ exploitation of previous data.
Crowdbreaks can serve as a framework to explore these challenges and find the right balance.

\subsubsection{Example use case}
The intensity, spread and effects of public opinion towards vaccination on social media and news sources has been explored in previous work~\cite{seeman2010assessing,Salathe2011a}.
Declines in vaccine confidence and boycotts of vaccination programs could sometimes be linked to disease outbreaks or set back efforts to eradicate certain diseases such as polio or measles~\cite{larson2011lessons,yahya2007polio}.
In particular, the potential benefits of real-time monitoring of vaccine sentiments as a tool for the improved planning of public health intervention programs has been highlighted~\cite{Larson2013,Pananos2017,Bahk2016}.\par
\begin{figure}[!ht]
\centering
\includegraphics{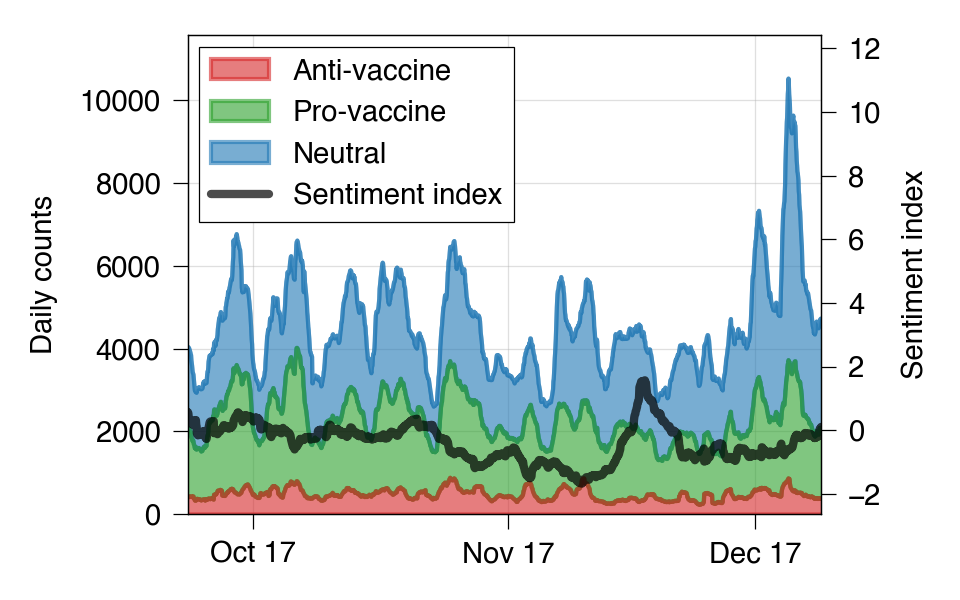}
  \caption{Real-time predictions of vaccine sentiments based on a Twitter data stream filtered by vaccine-related keywords. The colored values shown are stacked 1-day moving averages of tweet counts of the respective label class. The black curve denotes a sentiment index which reflects the ratio $r$ between counts of tweets labelled postive and negative in a one week moving average. The plotted line shows $(r-\mu)/\sigma$, in which $\mu$ and $\sigma$ denote mean and standard deviation, respectively.}
  \label{fig:fig4}
\end{figure}

Tracking of such sentiments towards vaccines is a primary use case of Crowdbreaks.
Figure~\ref{fig:fig4} shows real-time predictions based on a supervised bag-of-words fastText classifier~\cite{joulin2016bag}.
The predicted tweets were collected through the Twitter Streaming API using a list of vaccine-related keywords\footnote{The keywords include ``vaccine'', ``vaccination'', ``vaxxer'', ``vaxxed'', ``vaccinated'', ``vaccinating'', ``vacine''}.
The classifier was trained using publicly available data provided in recent work by Bauch et al.~\cite{Pananos2017}.
Please refer to their work for a detailed description of the collection and processing workflow of this data set.

\subsection{Technologies used}
Crowdbreaks uses a Python Flask API to interface between the components of the streaming pipeline and the user interface.
The streaming pipeline makes use of Redis for the message queuing of the processing queue as well as the priority queue (see figure~\ref{fig:fig1}).
Filtering and data processing, as well as NLP-related tasks are written in Python using the standard data analysis toolchain (numpy, scipy, nltk). 
Tweet objects are stored as flat files as well as in JSON format on Elasticsearch, which allows for an easier exploration and visualization of the data using Kibana.
The user interface is built using Ruby on Rails with a postgres database backend in order to store the annotations, as well as user-related data.

\section{Discussion \& Future work}
\label{sec:future_work}
Here we introduced Crowdbreaks, a tool allowing any researcher to start measurements of health and other trends in real-time from public social media content. 
By involving crowdworkers as well as the general public, we hope that these models will eventually improve to a level at which they can be incorporated into mathematical models in order to predict actual health indicators.
After proper validation and benchmarking, such models could eventually be used to improve public health decision-making, as well as risk assessments and disease forecasting.
In the case of disease prediction, the precise understanding of the content (e.g.\ whether a tweet just raises awareness vs. actually reporting an infection) is crucial for the robustness of the model.
As disease prediction solely from Twitter data remains to be a hard problem, previous work has suggested hybrid models between Twitter and less volatile data sources (such a Wikipedia page rate clicks) to be more robust \cite{McIver2014,Santillana2015}.
This may also serve as a future direction for disease prediction projects on Crowdbreaks.

\section{Acknowledgements}
We thank Sean Carroll, Yannis Jacquet, Djilani Kabaili  and S.P. Mohanty for valuable discussions and help regarding the technical aspects of this project. Thanks also to Chlo\'e All\'emann for comments on a draft of the paper.

\bibliography{refs}
\bibliographystyle{unsrt}

\end{document}